\documentclass{epl} 
\title{Emergence of rheological properties in lattice Boltzmann
  simulations of gyroid mesophases} \shorttitle{Rheol. properties of
  LB mesophases} 
\author{G. Giupponi\inst{1} \and J. Harting\inst{2}
  \and P.V. Coveney\inst{1}} \institute{
  \inst{1}Centre for Computational Science, Department of Chemistry, University College London - 20, Gordon Street, WC1H 0AJ, London, UK\\
  \inst{2}Institute for Computational Physics, University of Stuttgart
  - Pfaffenwaldring 27, 70569 Stuttgart, Germany }
\pacs{47.11.+j}{Computational methods in fluid dynamics}
\pacs{82.70.Uv}{Surfactants, micellar solutions, vesicles, lamellae,
  amphiphilic systems, (hydrophilic and hydrophobic interactions)}
\pacs{83.60.-a}{Rheology, Material behaviour}
\begin{document}

\maketitle

\begin{abstract}
  We use a lattice Boltzmann (LB) kinetic scheme for modelling
  amphiphilic fluids that correctly predicts rheological effects in
  flow.  No macroscopic parameters are included in the model.
  Instead, three-dimensional hydrodynamic and rheological effects are
  emergent from the underlying particulate conservation laws and
  interactions. We report evidence of shear thinning and viscoelastic
  flow for a self-assembled gyroid mesophase. This purely kinetic
  approach is of general importance for the modelling and simulation
  of complex fluid flows in situations when rheological properties
  cannot be predicted {\em a priori}.
\end{abstract}
\section{Introduction}
\begin{figure}
\onefigure[scale=0.28]{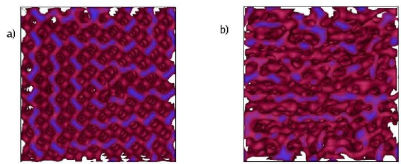}
\caption{a) High-density volume rendering of one of the two immiscible fluids (red) and interface between the two (blue) in the gyroid mesophase. b) Same as in a), but after 8000 simulation timesteps with an imposed steady shear ($U=0.1, \omega=0$ in eq.~(\ref{coseno})). The shear reduces the crystallinity and the material assumes more fluid-like properties. Model size is $64^3$}
\label{gyro}
\end{figure}
Lattice Boltzmann (LB) modelling schemes have emerged in the last
several years as a powerful approach for simulating the dynamics of a
variety of complex systems, from flow with suspended particles to
multiphase flow through porous media~\cite{bib:SucciBook}. Of great
interest is the extension of this method to the modelling of flow
properties of viscoelastic materials such as polymer melts and
amphiphilic fluids, where the dynamics of microscopic structures is
coupled to the flow.

Qian \and Deng~\cite{bib:Qian97} correctly described transverse wave
propagation using a lattice BGK model with an {\em ad hoc} modified
equilibrium distribution while Ispolatov \and
Grant~\cite{bib:Ispolatov02} obtained viscoelastic effects by adding a
force to represent memory effects in the LB equation.  However, both
approaches involve the inclusion of macroscopic parameters such as
Young's modulus~\cite{bib:Qian97} or elastic
coefficients~\cite{bib:Ispolatov02}. These methods cannot be regarded
as fully mesoscopic as at least one parameter is imposed on the basis
of macroscopic considerations. Giraud et al.~\cite{bib:Giraud98}
introduced a single fluid, two dimensional model to treat a
macroscopically viscoelastic fluid and later Lallemand et
al.~\cite{bib:Lallemand03} extended this model to three dimensions,
but to the best of our knowledge no generalization for multiphase flow
or flow with suspended particles has yet been implemented. A free
energy Ginzburg-Landau (GL) model can be defined to study rheological
properties of complex fluids\cite{bib:Patzold96, bib:Caputo02}. A
popular LB scheme based on the same GL approach proceeds by defining
an equilibrium distribution through the imposition of constraints on
macroscopic thermomechanical quantities such as the stress tensor.
Using such a scheme, Denninston et al.~\cite{bib:Denniston00} obtained
non-Newtonian flow behaviour (including shear thinning and banding)
using an LB algorithm for the hydrodynamics of liquid crystals in the
isotropic and nematic phases.  However, with such methods the dynamics
is not dictated by the mesoscopic processes; numerical
instability~\cite{bib:Kendon01} can make this approach unsuitable for
a fully mesoscopic description of the dynamics.

Gonz\'{a}lez and Coveney~\cite{bib:Segredo04}, using a fully
mesoscopic, kinetic approach which does not require the existence of a
thermodynamic potential obtained a self assembled gyroid mesophase in
the course of simulating an amphiphilic fluid formed by two immiscible
fluids and a surfactant species (fig.~\ref{gyro}). The term
amphiphilic fluid is used to describe a fluid in which at least one
species is made of surfactant molecules. Surfactants are molecules
comprised of a hydrophilic (water-loving) head group and a hydrophobic
(oil-loving) tail. An amphiphilic fluid may contain oil, water, or
both fluids in addition to surfactant. Such complex fluids can
self-assemble to form ordered structures such as lamellae, micelles,
sponge and liquid crystalline (cubic) phases showing pronounced
rheological properties\cite{bib:JonesBook}.
  
In this letter we use this purely kinetic LB method to model
complex flows whose rheological properties are emergent from the
mesoscopic kinetic processes without any imposed macroscopic
constraints\cite{bib:Chen99}. In particular, we show evidence of the
appearance of rheological effects, such as shear thinning and
viscoelasticity, for a self-assembled gyroid liquid crystalline cubic
mesophase.

\section{The model}

A standard LB system involving multiple species is usually represented
by a set of equations~\cite{bib:Higuera89}
\begin{equation}
\label{LBeqs}
\begin{array}{cc}
n_i^{\alpha}({\bf x}+{\bf c}_i, t+1) - n_i^{\alpha}({\bf x},t) =  -\frac{1}{\tau_{\alpha}}(n_i^{\alpha}({\bf x},t) - n_i^{\alpha
 eq}({\bf x},t)) , &  i= 0,1,\dots,b\mbox{ ,}
\end{array}
\end{equation}
where $n_i^{\alpha}({\bf x},t)$ is the single-particle distribution
function, indicating the density of species $\alpha$ (for example,
oil, water or amphiphile), having velocity ${\bf c}_i$, at site ${\bf
  x}$ on a D-dimensional lattice of coordination number $b$, at
time-step $t$. The collision operator $\Omega_i^{\alpha}$ represents
the change in the single-particle distribution function due to the
collisions.  We choose a single relaxation time $\tau_{\alpha}$, `BGK' form~\cite{bib:Bhatnagar54} for the collision operator
%
In the limit of low Mach numbers, the LB equations correspond to a
solution of the Navier-Stokes equation for isothermal,
quasi-incompressible fluid flow whose implementation can efficiently
exploit parallel computers, as the dynamics at a point requires only
information about quantities at nearest neighbour lattice sites.  The
local equilibrium distribution $n_i^{\alpha eq}$ plays a fundamental
role in the dynamics of the system as shown by eq.~(\ref{LBeqs}). In
this study, we use a purely kinetic approach, for which the local
equilibrium distribution $n_i^{\alpha eq}({\bf x},t)$ is derived by
imposing certain restrictions on the microscopic processes, such as
explicit mass and total momentum conservation~\cite{bib:Chen91}
\begin{equation}
\label{Equil}
n_i^{\alpha eq} = \zeta_in^{\alpha}\left[1+
\frac{{\bf c}_i \cdot {\bf u}}{c_s^2} +\frac{({\bf c}_i \cdot {\bf u})^2}{2c_s^4}
-\frac{u^2}{2c_s^2}+\frac{({\bf c}_i \cdot {\bf u})^3}{6c_s^6}
-\frac{u^2({\bf c}_i \cdot {\bf u})}{2c_s^4}\right] \mbox{ ,}
\end{equation}
where ${\bf u} = {\bf u}({\bf x},t)$ is the macroscopic bulk velocity
of the fluid, defined as $n^{\alpha}({\bf x},t){\bf u}^{\alpha}
\equiv \sum_i n_i^{\alpha}({\bf x},t){\bf c}_i$, $\zeta_i$ are the
coefficients resulting from the velocity space discretization and
$c_s$ is the speed of sound, both of which are determined by the
choice of the lattice, which is D3Q19 in our implementation.
Immiscibility of species $\alpha$ is introduced in the model following
Shan and Chen~\cite{bib:Shan93, bib:Shan94}. Only nearest neighbour
interactions among the immiscible species are considered.  These
interactions are modelled as a self-consistently generated mean field
body force
\begin{equation}
{\bf F}^{\alpha}({\bf x},t) \equiv -\psi^{\alpha}({\bf x},t)\sum_{\bf \bar{\alpha}}g_{\alpha \bar{\alpha}}\sum_{\bf x^{\prime}}\psi^{\bar{\alpha}}({\bf x^{\prime}},t)({\bf x^{\prime}}-{\bf x})\mbox{ ,}
\end{equation}
where $\psi^{\alpha}({\bf x},t)$ is the so-called effective mass,
which can have a general form for modelling various types of fluids
(we use $\psi^{\alpha} = (1 - e^{-n^{\alpha}}$)\cite{bib:Shan93}), and
$g_{\bar{\alpha}\alpha}$ is a force coupling constant whose magnitude
controls the strength of the interaction between components $\alpha$,
$\bar{\alpha}$ and is set positive to mimic repulsion. The dynamical
effect of the force is realized in the BGK collision operator in
eq.~(\ref{Omega}) by adding to the velocity ${\bf u}$ in the
equilibrium distribution of eq.~(\ref{Equil}) an increment

\begin{equation}
\delta{\bf u}^{\alpha} = \frac{\tau^{\alpha}{\bf F}^{\alpha}}{n^{\alpha}}\mbox{ .}
\end{equation}

As described above, an amphiphile usually possesses two different
fragments, each having an affinity for one of the two immiscible
components. The addition of an amphiphile is implemented as in
~\cite{bib:Chen99}. An average dipole vector ${\bf d}({\bf x},t)$ is
introduced at each site ${\bf x}$ to represent the orientation of any
amphiphile present there. The direction of this dipole vector is
allowed to vary continuously and no information is specified for each
velocity ${\bf c}_i$, for reasons of computational efficiency and
simplicity. Full details of the model can be found
in~\cite{bib:Chen99} and~\cite{bib:Nekovee00}.

\section{Simulation details} 

We use LB3D\cite{bib:Harting05}, a highly scalable parallel LB code,
to implement the model. A single simulation of a $64^3$ model, i.e. a
point in fig.\ref{fig1} below, needs around 300 Mbytes of RAM and
takes about 50 CPU hours to complete on a single processor machine.
The very good scaling of our LB3D code permits us to run all our
simulations on multiprocessor machines and computational grids in
order to reduce the length of data collection to a few weeks. The
simulation parameters are those used in~\cite{bib:Segredo04} that lead
to the formation of gyroid mesophases. We use a $64^3$ lattice size
with periodic boundary conditions. The simulations were all started
from a checkpointed configuration of a $200000$ timesteps equilibrated
gyroid mesophase\cite{bib:Harting05}.  A single unit cell of the
gyroid minimal surface is of the order of $5-8$ lattice lengths in its
linear dimensions. This implies that hundreds of unit cells are
present in a $64^3$ sample. Therefore, finite size effects are not
expected to affect the qualitative outcome of the simulations.
Moreover, some simulations with a $128^3$ lattice have been run to
confirm the results obtained and the $128^3$ simulations reported
in~\cite{bib:Segredo05} are consistent with the results shown here.
Nevertheless, such lattice sizes oblige us to study perfect gyroids;
much larger system sizes (and computational cost) would be necessary
in order to investigate the role of domains and their associated
defects\cite{bib:defect-paper}. We leave this for future work.

In order to inspect the rheological behaviour of gyroid mesophases, we
have implemented Lees-Edwards boundary conditions, which reduce finite
size effects if compared to moving solid walls~\cite{bib:Lees72}. This
computationally convenient method imposes new positions and velocities
on particles leaving the simulation box in the direction perpendicular
to the imposed shear strain while leaving the other coordinates
unchanged. Choosing $z$ as the direction of shear and $x$ as the
direction of the velocity gradient, we have

\begin{eqnarray}
z^{\prime} \equiv \left\{
\begin{array}{ll}
(z+\Delta_z) \mbox{ mod $N_z$}&,x > N_x  \\
z \mbox{ mod $N_z$}&,0\le x \le N_x   \\
(z-\Delta_z) \mbox{ mod $N_z$}&,x < 0    \\
\end{array} \right.
&
u_z^{\prime} \equiv \left\{
\begin{array}{ll}
u_z+U&,x > N_x  \\
u_z&,0\le x \le N_x   \\
u_z-U&,x < 0    \\
\end{array} \right.\mbox{ ,}
\end{eqnarray}
where $\Delta_z \equiv U\Delta t$, U is the shear velocity, $u_z$ is
the $z-$component of ${\bf u}$ and $N_{x(z)}$ is the system length in
the $x(z)$ direction. We also use an interpolation scheme suggested by
Wagner \and Pagonabarraga~\cite{bib:Wagner01} as $\Delta_z$ is not
generally a multiple of the lattice site. Cates et
al.~\cite{bib:Cates04}, found pronounced artefacts (lock-ins) in
simulations of 2D sheared binary mixtures with multiple Lees-Edwards
planes. In our own work, we have not seen any of these artefacts, even
in the longest simulations performed, and a linear velocity profile is
obtained at steady state.

Consistent with the hypothesis of the LB model, we set the maximum
shear velocity to $U=0.1$ lattice units. This results in a maximum
shear rate $\dot{\gamma}_{xz}=\frac{2\times0.1}{64}=3.2\times10^{-3}$
in lattice units.  Simulations are run for $T=10000$ timesteps in the
case of steady shear. Steady state is reached in the first $1000$
timesteps, and the relevant component of the stress tensor is averaged
over the last $3000$ timesteps. Some measurements were repeated by
doubling the simulation time to a total of $T=20000$ timesteps but no
significant differences were found. For oscillatory shear, we set
\begin{equation}
\label{coseno}
U(t) = U\cos(\omega t)\mbox{ ,}
\end{equation}
where $\omega/2\pi$ is the frequency of oscillation and $U=0.05$
lattice units.  These simulations were run for at least three complete
oscillations, $t=5000-100000$ timesteps.

\section{Results} 

\begin{figure}
\twofigures[scale=0.25]{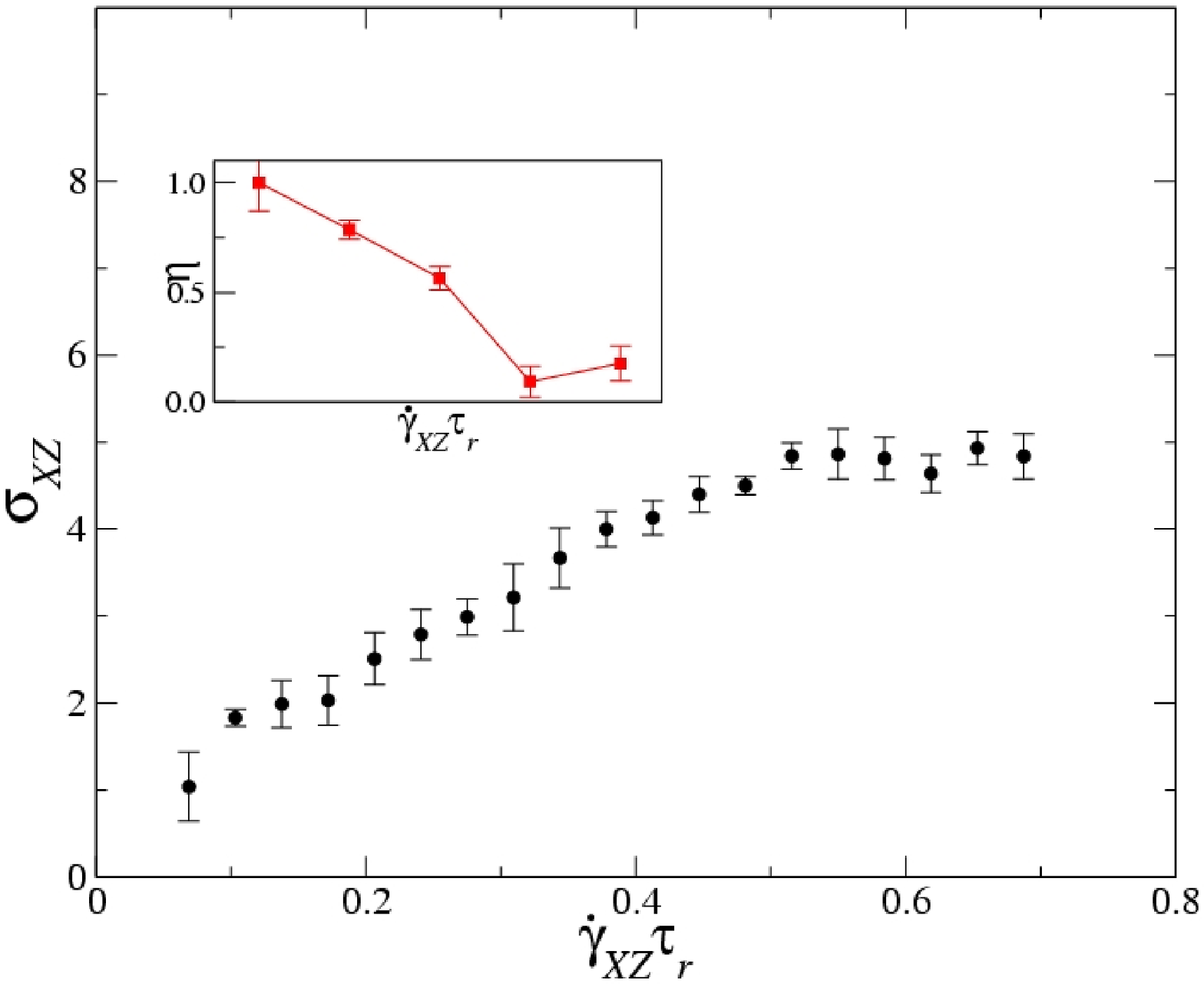}{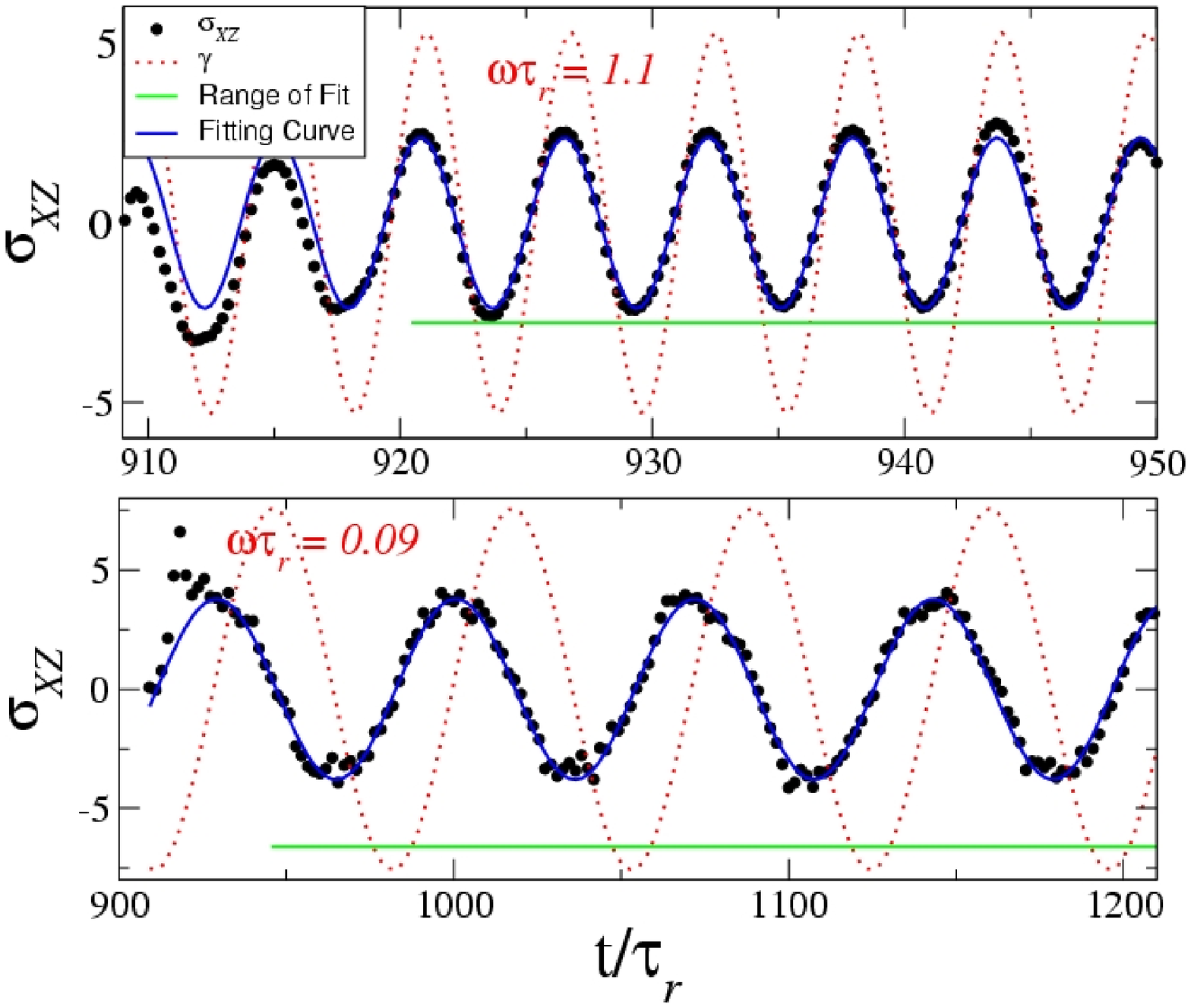}
\caption{Plot of $\sigma_{xz}$ (dimensionless units) vs. shear rate
$\dot{\gamma}_{xz}\tau_r$ (Weissenberg number) for an LB simulation of a $64^3$
gyroid mesophase. $\tau_r$ is the stress relaxation time; see text for discussion. The inset shows the percentage drop of the viscosity as the shear rate $\dot{\gamma}_{xz}\tau_r$ increases (shear-thinning).}
\label{fig1}
\caption{Plot of $\sigma_{xz}$ (black dots) vs. time for oscillatory shear 
  ($U=0.05$ and $\omega = 0.0004, 0.005$ in eq.~(\ref{coseno}))
  applied to $64^3$ lattice-Boltzmann gyroid mesophase. All quantities are
dimensionless. After a brief
  transient, $\sigma_{xz}$ oscillates with the imposed frequency
  $\omega\tau_r$. The strain (dashed line) is plotted (scaled) as a
  guide to the eye. The continuous line is the best fit line for the
  $\sigma_{xz}$ data. Shear moduli $G^{\prime}(\omega\tau_r)$,
  $G^{\prime\prime}(\omega\tau_r)$ are calculated as in
  eq.~(\ref{calmod}) over a time interval shown by the continuous
  horizontal line.}
\label{fig2}
\end{figure}

In order to investigate rheological properties of the system, we
perform simulations of flow under shear, mimicking a rheometer. We set
$U = n\times0.005$, $n=2\dots 21$ and $\omega = 0$ to impose a
stationary Couette flow. Once steady state flow is reached, we collect
the relevant component of the pressure tensor $\sigma_{xz}$.
According to Newton's law for viscous flow of a liquid we have

\begin{equation}
\sigma_{xz} = 2\eta\dot{\gamma}_{xz}\mbox{ ,}
\label{liquid}
\end{equation} 
where $\eta$ is the viscosity of the liquid and $\dot{\gamma}_{xz} =
\frac{(\partial_{x}{u_z} + \partial_{z}{u_x})}{2}$. In fig.~\ref{fig1}
we plot $\sigma_{xz}$ against the imposed non-dimensionalized (see
text below) shear rate $\dot{\gamma}_{xz}\tau_r$ (Weissenberg number).
We note that the slope of the curve changes as these shear rate
increases. This indicates that the viscosity $\eta$ depends on the
shear rate $\dot{\gamma}_{xz}$, $\eta = \eta(\dot{\gamma}_{xz})$. We
therefore conclude that our model exhibits non-Newtonian flow
behaviour. Complex fluids such as amphiphilic mixtures are well known
to exhibit such behaviour experimentally~\cite{bib:JonesBook}. In the
inset we calculate the percentage drop of $\eta$ by averaging the slope
of the curve over subsets of four data points.  We note that the
viscosity decreases as the shear rate increases. This behaviour is
referred to as ``shear thinning''~\cite{bib:JonesBook}.  In general,
fluids exhibit macroscopic non-Newtonian properties because of
underlying complex mesoscopic interactions and leading to changes in
their microstructural properties. In our case, the highly ordered
structure of the gyroid mesophase is responsible for this rheological
effect, and our LB model correctly captures it. Therefore, considering
the underlying physics of mesophases, this simulation result is not
only justified but also expected. We note that the predictions of the
model could be directly verified, but no experimental evidence for
gyroid rheology is currently available in the literature.

\begin{figure}
  \twofigures[scale=0.29]{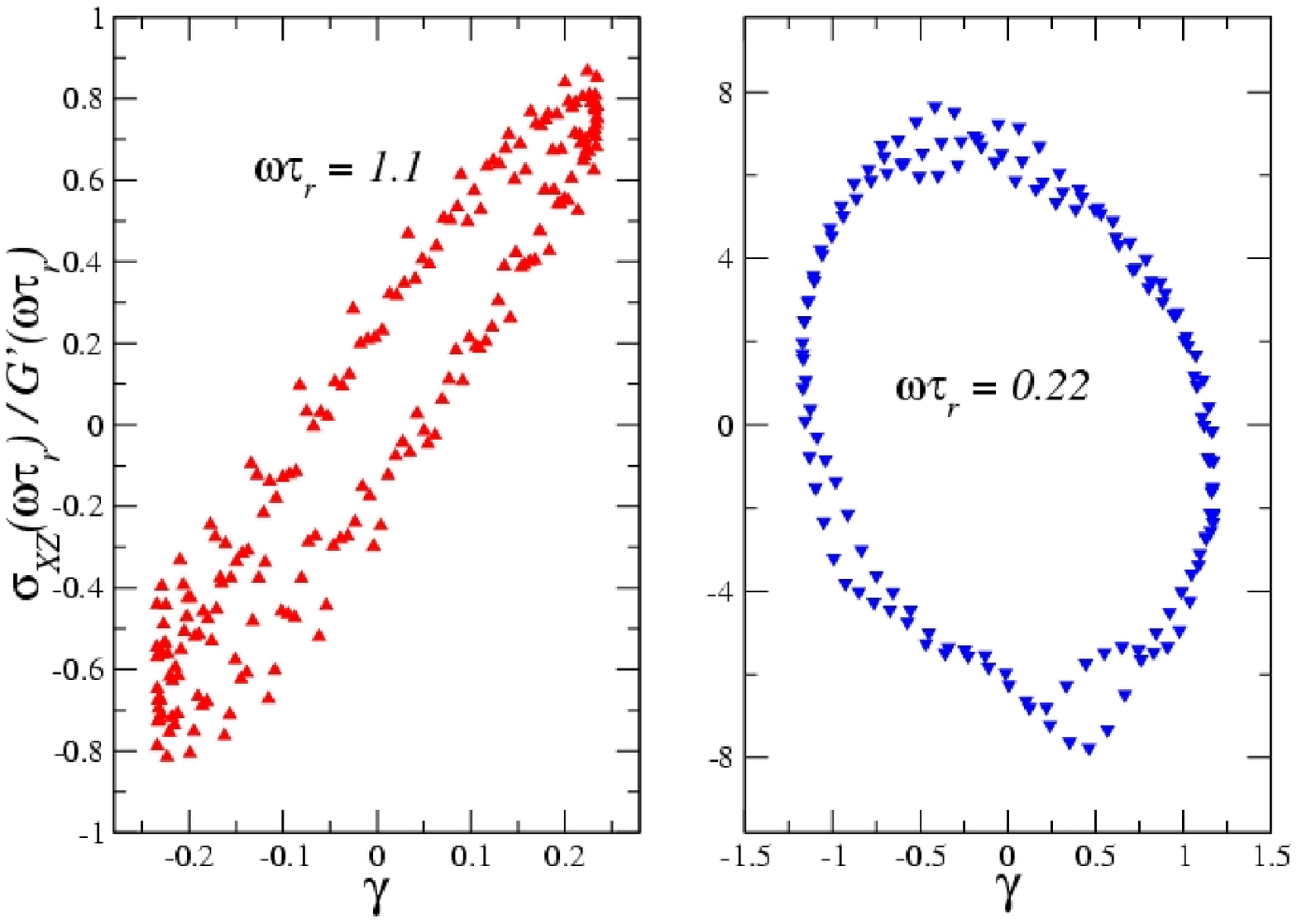}{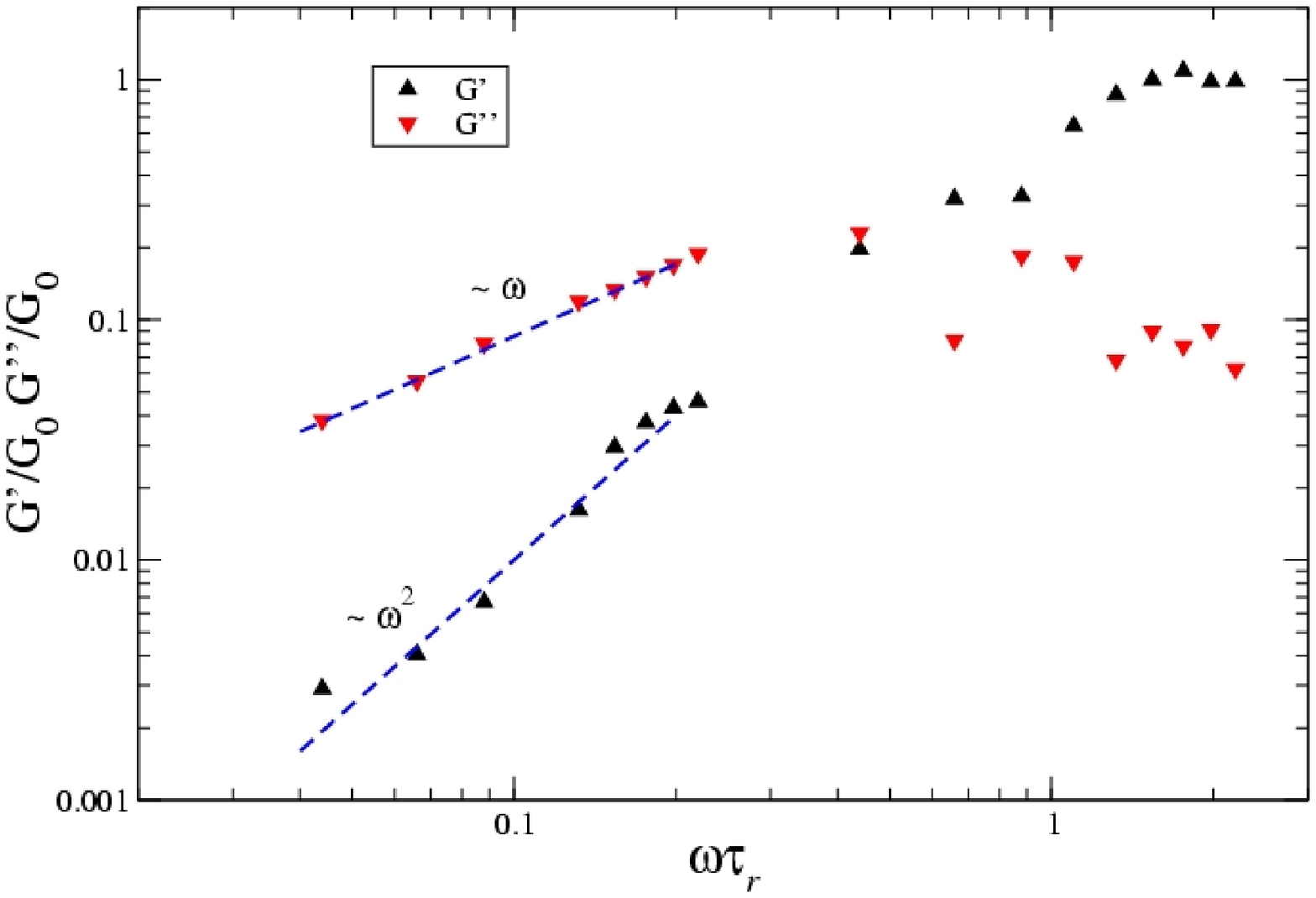}
\caption{Lissajous plots of $\sigma_{xz}$ (in lattice units) vs. dimensionless strain $\gamma_{xz}$. In the left(right) panel, points tend to form a straight line (ellipse, respectively). This indicates that the pressure $\sigma_{xz}$ is in(out of) phase with respect to the strain $\gamma_{xz}$, as typical for solid(liquid)-like phases. The overall behaviour is typical of viscoelastic fluids.}
\label{fig3}
\caption{Plots of shear moduli $G^{\prime}(\omega\tau_r)$ (regular triangles), $G^{\prime\prime}(\omega\tau_r)$ (inverted triangles) calculated for all frequencies used in our simulations. We note that a crossover between liquid and solid-like behaviour occurs around $\omega=0.5$. Theoretically, $G^{\prime}\sim \omega^2$, $G^{\prime\prime} \sim \omega$ as $\omega$ goes to $0$. Our data agree with this prediction as shown by the dashed lines in the plot. All quantities are dimensionless ($G_0$ is the plateau modulus for $G'$).}
\label{fig4}
\end{figure}

For oscillatory shear, in eq.~(\ref{coseno}) we set $U=0.05$ lattice
units and we span a range of two decades of frequencies by varying
$\omega$ between $\omega = 0.0001$ and $\omega = 0.01$ lattice units.
For a viscoelastic medium
\begin{equation}
\label{moduli}
\sigma_{xz} = \gamma_0 [G^{\prime}(\omega)\sin(\omega t) + G^{\prime\prime}(\omega)\cos(\omega t)]\mbox{ ,}
\end{equation} 
where $\gamma_0 = \frac{2U}{64\omega}$ is the imposed shear strain and
$G^{\prime}(\omega)$, $G^{\prime\prime}(\omega)$ are, respectively,
the storage and loss moduli which respectively measure the elastic and
viscous response at any given frequency~\cite{bib:JonesBook}. In
fig.~\ref{fig2} we show the time dependence of $\sigma_{xz}$ for two
simulations with different values of $\omega$. After a brief
transient, $\sigma_{xz}$ begins to oscillate as predicted by
eq.~(\ref{moduli}) with the imposed frequency $\omega$ and an
amplitude $\sigma_0$ and shift $\phi$ that depend on the storage and
loss moduli

\begin{eqnarray}
\label{calmod}
G^{\prime} = \frac{\sigma_0 \cos(\phi)}{\gamma}
&\mbox{   } &
G^{\prime\prime} = \frac{\sigma_0 \sin(\phi)}{\gamma}
\mbox{ ,}
\end{eqnarray}
where $\sigma_0$ and $\phi$ are the fitted values for the amplitude
and phase shift. We derive $G^{\prime}(\omega)$ and
$G^{\prime\prime}(\omega)$ for the two different values of $\omega =
0.005, 0.0004$ by fitting $\sigma_0$ and $\phi$ with the standard
Levenberg-Marquardt algorithm over at least three decades (see caption
for details). In fig.~\ref{fig3}, we plot $\sigma_{xz}$ against the
shear strain at different times (Lissajous plots) for the two
frequencies of fig.\ref{fig2}. We note that for the higher frequency
(left panel), $\sigma_{xz}$ is in phase with the strain, indicating
that the gyroid mesophase exhibits a solid-like response at short
timescales. For the smaller frequency, the phase shift
is almost $\pi/2$, typical of a liquid-like response as dictated by
eq.~(\ref{liquid}). In fig.~\ref{fig4}, we plot $G^{\prime}(\omega\tau_r)$
and $G^{\prime\prime}(\omega\tau_r)$ for the different simulation values of
$\omega$. In order to capture the relevant scales in the model, all
quantities are dimensionless. In particular, the frequency $\omega$
has been multiplied by the characteristic stress relaxation time,
$\tau_r \sim 220$ simulation timesteps, calculated by fitting the
exponential decay of $\sigma_{xz}$ when shear is removed.
We note that a crossover between $G^{\prime}(\omega\tau_r)$ and
$G^{\prime\prime}(\omega\tau_r)$ occurs at $\omega\tau_r \sim 0.05$.
This signals a transition between solid and liquid-like behaviour and
indicates that our LB model is capable of predicting viscoelastic
flow; see also fig.~(\ref{gyro}) and caption. We note that a
dimensionless crossover point of $\omega\tau_r \sim 1$ consistently
links rheological properties of the mixture with the relaxation time
of the gyroid mesophase. This is encouraging for the use of this
simulation model to inspect the correlation between macroscopic
dynamics and mesoscopic structure. Indeed in the case of linear
viscoelasticity, $G^{\prime} \sim \omega^2$, $G^{\prime\prime} \sim
\omega$ as $\omega$ goes to $0$ for any fluids.  In fig.~\ref{fig4} we
plot lines $\sim \omega$ and $\sim\omega^2$ as a guide for the eye;
our data show excellent quantitative agreement with theory. We
interpret the elastic response component as due to the stress response
to mechanical perturbations of the long-range ordered equilibrium
structure of the liquid gyroid crystalline mesophase.  We also tried
to fit the data in fig.\ref{fig4} to a simple single relaxation time
(Maxwell) model of viscoelasticity but the fit is very poor. This
suggests that there are several relaxation times present, as is to be
expected here since the viscoelasticity arises from complex mesoscale
structure. We note that the existence of a spectrum of relaxation
times is consistent with the stretched-exponential behaviour of domain
self-assembly in amphiphilic fluid systems found
in~\cite{bib:Segredo04}.

\section{Conclusions}

In this letter we use a purely kinetic LB model that leads to the
emergence of non-trivial rheological properties. Non-Newtonian
behaviour, in this case a decrease of viscosity with shear rate, has
been shown for an initially self-assembled gyroid liquid crystalline
mesophase. In addition, linear viscoelastic effects in the system are
manifest in the simulations. It is notable that our model correctly
predicts the theoretical limits for the moduli $G^{\prime}(\omega)$,
$G^{\prime\prime}(\omega)$ as $\omega$ goes to $0$ as well as a
crossover in $G^{\prime}(\omega)$, $G^{\prime\prime}(\omega)$ at
higher $\omega$. We note that, unlike previous approaches, this model
does not require any assumptions at the macroscopic level, that is it
provides a purely kinetic-theoretical approach to the description of
complex fluids.  This model can therefore help in understanding
complex flows, such as flow of viscoelastic liquids in porous media,
colloidal fluids and polymer melts. Work is in progress to assess the
generality of this approach within our lattice-Boltzmann model of
amphiphilic fluids. In particular, such simulations should enable us to
investigate the link between mesoscopic structure and macroscopic
dynamics, for example by correlating rheological relaxation times to
mesoscale relaxation processes within the amphiphilic fluid.

\acknowledgments We would like to thank D.M.A. Buzza for a critical
reading of this manuscript, as well as J. Chin and N. Gonz\'alez-Segredo for
helpful discussions. This research is funded by EPSRC under the
RealityGrid grant GR/R67699 which also provided access to UK national
supercomputing facilities at CSAR and HCPx. Access to the US TeraGrid
and Lemieux at Pittsburgh Supercomputing centre was provided through
the National Science Foundation (USA) NRAC and PACS grants MCA04N014
and ASC030006P respectively. Intercontinental data tranfer was
effected via UKLight within EPSRC grant GR/T04465. G.G. would like to
acknowledge the support of the European Commission's Research
Infrastructure Activity, contract number 506079 (HPC Europa).

\end{document}